\documentclass[aps,showpacs,twocolumn,preprintnumbers]{revtex4}
\usepackage{bm}
\begin{document}
%\input psfig

%\preprint{{DOE/ER/XXXXX-XXX}\cr{UMPP\#05-XXX}}

\count255=\time\divide\count255 by 60 \xdef\hourmin{\number\count255}
  \multiply\count255 by-60\advance\count255 by\time
 \xdef\hourmin{\hourmin:\ifnum\count255<10 0\fi\the\count255}

\newcommand{\xbf}[1]{\mbox{\boldmath $ #1 $}}

\newcommand{\sixj}[6]{\mbox{$\left\{ \begin{array}{ccc} {#1} & {#2} &
{#3} \\ {#4} & {#5} & {#6} \end{array} \right\}$}}

\newcommand{\threej}[6]{\mbox{$\left( \begin{array}{ccc} {#1} & {#2} &
{#3} \\ {#4} & {#5} & {#6} \end{array} \right)$}}

\newcommand{\clebsch}[6]{\mbox{$\left( \begin{array}{cc|c} {#1} & {#2} &
{#3} \\ {#4} & {#5} & {#6} \end{array} \right)$}}

\newcommand{\iso}[6]{\mbox{$\left( \begin{array}{cc||c} {#1} & {#2} &
{#3} \\ {#4} & {#5} & {#6} \end{array} \right)$}}

\title{Decoupling Spurious Baryon States in the $1/N_c$  Expansion of QCD}

\author{Thomas D. Cohen}
\email{cohen@physics.umd.edu}

\affiliation{Department of Physics, University of Maryland, College
Park, MD 20742-4111}

\author{Richard F. Lebed}
\email{Richard.Lebed@asu.edu}

\affiliation{Department of Physics and Astronomy, Arizona State
University, Tempe, AZ 85287-1504}

%\date{\hourmin, \today}
\date{April, 2006}

\begin{abstract}
We identify ``spurious'' states in meson-baryon scattering, those that
appear in QCD for $N_c \! > \! 3$ but decouple for $N_c \! = \! 3$.
The key observation is that the relevant flavor SU(3) Clebsch-Gordan
coefficients contain factors of $1 \! - \! 3/N_c$.  We show that this
method works even if SU(3) is badly broken.  We also observe that
resonant scattering poles lying outside naive quark model multiplets
are not necessarily large $N_c$ artifacts, and can survive via
configuration mixing at $N_c \! = \! 3$.
\end{abstract}

\pacs{11.15.Pg, 14.20.Gk, 14.20.Jn}
%11.15.Pg   Expansions for large numbers of components (e.g.,
%           1/Nc expansions)
%14.20.Gk   Baryon resonances with S=0
%14.20.Jn   Hyperons

\maketitle

\section{Introduction}

A large body of literature convincingly demonstrates that the $1/N_c$
expansion about the limit $N_c \! \to \! \infty$ provides a fertile
harvest of qualitative and semi-quantitative information about QCD\@.
However, the study of baryons in large $N_c$ QCD presents a formidable
complication absent from the meson sector, namely, the emergence of
{\em spurious states\/}, which appear in $N_c \! > \!  3$ worlds but
have no analog in the physical $N_c \! = \!  3$ world.  Clearly, the
measurable properties of states that exist at $N_c \!  = \! 3$ must
decouple from the spurious states as $N_c \! \rightarrow \! 3$ in a
systematic $1/N_c$ expansion.  In this context, decoupling refers both
to static and dynamical properties: The matrix elements of neither
type of observable for physical states at $N_c \!  = \! 3$ may be
permitted to depend on couplings to states that do not exist in the
$N_c \! = \! 3$ world.  To keep the discussion concise, we denote
``physical'' states as ones whose analogs survive at $N_c \!  = \! 3$
and ``spurious'' as ones that do not.  One can imagine starting with a
world of large (but finite) $N_c$; then the spurious states are
allowed and couple to physical states, in the sense that they can be
reached from the physical states via the emission or absorption of
some number of light mesons.  As $N_c$ is reduced towards 3, such
couplings must diminish; and eventually as one reaches the world of
$N_c \! = \! 3$, all couplings between the physical and spurious
states must vanish.  This paper explores how this decoupling comes
about.

In fact, this is a multifaceted and complicated problem.  Before
proceeding any further we must clarify the precise attributes of
spurious states.  We mean here states that, by virtue of their quantum
numbers, {\em cannot\/} exist in $N_c \! = \! 3$ QCD\@.  We
distinguish such states from those that merely fail to exist in the
most naive quark model (such as pentaquarks).  To make this
distinction clear, let us begin by considering a world with two light
flavors and large (but finite) $N_c$, and focus on the ground-state
band of states.  As is well known~\cite{DJM1} these states have $I \!
= \! J$, and their low-lying excitation energies [including a common
mass $M_0 \! = \!  O(N_c^1)$] are given by
\begin{equation}
M_J = M_0 + \frac{J(J+1)}{2 {\cal I}} + O (1/N_c^2) \ ,
\label{mj}
\end{equation}
where ${\cal I} \! = \! O(N_c^1)$ is a moment of inertia parameter.
As $N_c \! \rightarrow \! \infty$ the system is represented by a
contracted SU(4) symmetry~\cite{DJM1}, and one finds an arbitrarily
large number of states degenerate in mass to within $O(1/N_c)$.  Note
that for any nonzero $m_\pi \! = \! O(N_c^0)$, states with $I \! = \!
J \! = \!  O(N_c^0)$ are stable since there is no phase space for
decay.  Of course, in the real world approximate chiral symmetry
ensures an anomalously small $m_\pi$.  In the exact chiral limit of
$m_\pi \! = \! 0$ the phase space for the decay of such states is not
zero.  However, even in this limit, the states with $I \! = \! J \! =
\!  O(N_c^0)$ remain narrow, with $O(1/N_c)$ widths.

For finite $N_c$ the preceding picture for ground-state baryons
clearly breaks down, and Eq.~(\ref{mj}) breaks down with it.  In the
context of a naive quark model with all of the quarks in s waves,
$J_{\rm max} \! = \! \frac{N_c}{2}$.  Moreover, states with $I
\! > \!  \frac{N_c}{2}$ are impossible in the context of a naive quark
model allowing only minimal quark content.  Thus, states with $I \! =
\! J \! > \! \frac{N_c}{2}$ might then be labeled spurious.  However,
this argument is based upon a naive quark model description and not
upon full QCD.  So, what happens in QCD for large but finite $N_c$?
The answer is that no one knows in detail; one would have to be able
to solve QCD completely to find out.  But there does exist a
qualitative idea that is phenomenology consistent.  One notes that the
level spacings from Eq.~(\ref{mj}) scale as
\begin{equation}
M_{J+1}-M_J = \frac{J +1}{\cal I} = O \left( J/N_c \right) \ .
\end{equation}
Thus, for $J \! < \! O(N_c^1)$ the level spacing is smaller than
$O(N_c^0)$.  The phase space for the decay of the $J \! + \! 1$ state
into the $J$ state plus a $\pi$ is small [either it is below $m_\pi$
and hence zero, or if one works in the chiral limit it is driven by
the $O(1/N_c)$ mass splitting], and hence the particle masses are well
defined.  Now suppose one took this result seriously for larger values
of $J$; it becomes clear that something must break down.  For example,
as $J$ increases towards $O(N_c^1)$ the width of the state becomes
$O(N_c^0)$, and one can no longer speak of narrow states given by
Eq.~(\ref{mj}).  In one qualitative sense this is similar to the fate
of the naive quark model at large but finite $N_c$: In both cases
Eq.~(\ref{mj}) breaks down for $I \! = \! J \! = \! O(N_c^1)$.
However, one also finds an important distinction.  In the naive quark
model (with minimal quark content), the states with $I \! = \! J
\! > \! \frac{N_c}{2}$ are spurious.  In a more general picture of
large $N_c$ QCD, the breakdown of Eq.~(\ref{mj}) is gradual as $J$
increases, and no point occurs at which one can identify the states as
spurious.  QCD with a large but finite value of $N_c$ allows baryon
number unity states with $I \! = \! J$ quantum numbers for all
half-integral values of $I \!  = \! J$.  These states may not all be
narrow or indeed may not even be resonant, but they are not forbidden.
The possibility of states with $I \! = \! J \! > \! \frac 3 2$
existing in the context of large $N_c$ models such as the Skyrme model
has long been known~\cite{CG}.  Thus, in the language used here these
states are not spurious.  We denote such states as ``exotic'', since
they are states beyond those accessible in the naive quark model with
minimal quark content.

The example considered above illustrates an important issue, that the
worrisome states---regardless of whether labeled ``spurious'' or
``exotic''---are precisely those with large $I$ and $J$.  Note that
these states only connect to the nucleon via the emission of a large
number of mesons, which is necessary to support a large isospin value.
Thus, if one restricts to baryon resonances obtainable in
meson-nucleon scattering, states with large $I$ or $J$ are not a
concern.  Scattering states are useful to examine not only because a
large fraction of the known resonances are in fact observed in these
channels, but also because there has been considerable theoretical
progress in understanding these processes at large
$N_c$~\cite{CL1st,CLcompat,CLconfig,CL1N,CLphoto,CLSU3,CLSU3penta,CL70,ItJt}.

Although the situation is quite simple for a 2-flavor world, the
generalization to 3 flavors is drastically different.  The allowable
2-flavor representations in an $N_c \! = \! 3$ world are equivalent to
those found in a large $N_c$ world.  However, {\em none\/} of the
3-flavor multiplets seen at $N_c \! = \!  3$ exist in a large $N_c$
world, and conversely, {\em all\/} of the 3-flavor multiplets for
large $N_c$ contains states that are spurious at $N_c \! = \! 3$.  The
reason for this disparity is clear: In the exact SU(3) limit $I$-spin,
$U$-spin, and $V$-spin all exist on equal footing.  In the SU(2)
flavor space one may consider only representations where the isospin
is small, i.e., $I \! = \!  O(N_c^0)$ as opposed to $I \! = \!
O(N_c^1)$.  However, in SU(3) flavor if one chooses representations
with $I \! = \! O(N_c^0)$, then necessarily $U \! = \! O(N_c^1)$ and
$V \! = \!  O(N_c^1)$.  Thus, for example, the spin-$\frac 1 2$
baryons of the ground-state band form the nucleon isodoublet for 2
flavors at any $N_c$, but for 3 flavors occupy a $(p,q)= \left[ 1 ,
\frac 1 2 (N_c \! - \! 1) \right] \! \equiv$ ``{\bf 8}''
representation.  For $N_c \! = \!  3$ this representation is the
familiar octet, but as $N_c \! \rightarrow \! \infty$ it grows
arbitrarily large.  When $N_c$ is large but finite most of the states
in this representation are spurious, having no $N_c \! = \! 3$ analog.
Even the ground-state baryon representations include spurious states.

As noted above, we wish to access excited baryons via meson-baryon
scattering on a target baryon in its ground state.  Inasmuch as one
can study this scattering using an arbitrary value of $N_c$, it must
be possible to use the continuous set of results connecting the limit
$1/N_c \! = \! 0$ to the physical point $1/N_c \!  = \! 1/3$.  In the
latter case, all spurious states must therefore decouple; the theory
must forbid access to these states at $N_c \! = \! 3$.  Moreover, this
decoupling must be smooth; such couplings must diminish gradually and
vanish as $N_c$ reaches 3.

The goal of the present paper is to understand how this decoupling
occurs.  It is a critical task since $1/N_c$ is not very small for
$N_c \! = \! 3$; in order to obtain more than semi-quantitative
predictions it is essential to be able to handle the $1/N_c$
corrections.  The coupling to spurious states clearly represents a key
class of $1/N_c$ corrections.  Moreover, the nature of these
corrections is rather special: Unlike typical power series
corrections, which become progressively less significant at higher
orders, the couplings to spurious states impose selection rules and
completely forbid access to certain states.

\section{Types of $1/N_c$ Corrections in Meson-Baryon Scattering}
\label{idea}

Since large $N_c$ QCD has a smooth $1/N_c \! \to \! 0$ limit that
supports a large number of spurious states, it must be true that at
$N_c \! = \! 3$ the sum of the various $1/N_c$ corrections conspire to
cancel exactly the leading-order result for all couplings between
physical and spurious states, while retaining all physical states.  At
first sight it may seem daunting to show how such a conspiracy can
come about.  Clearly, it is necessary to understand first the possible
sources of $1/N_c$ corrections to the theory.

It is well known that the scattering amplitudes are connected via
linear relations at leading order in $1/N_c$.  These relations follow
from a master equation in which the physical amplitudes are linear
superpositions of various ``reduced'' amplitudes~\cite{MP}.  In light
of these relations, it is easy to see that $1/N_c$ corrections to the
scattering amplitudes may grouped into three classes:

$i)$ There are $O(1/N_c)$ corrections to the leading-order
[$O(N_c^0)$] reduced amplitudes $\tau$; one can show~\cite{ItJt}
that all such amplitudes multiply structures obeying the simple
$t$-channel rules $I_t \! = \! J_t$ and $Y_t \! = \!  0$.  The
master expression obtained in Ref.~\cite{ItJt} is:\begin{widetext}
\begin{eqnarray}
\lefteqn{S_{L L^\prime S_{B^{\vphantom\prime}} S_{B^\prime} J_t J_{tz}
R_t \gamma^{\vphantom\prime}_t \gamma^\prime_t I_t Y_t} =
\delta_{J^{\vphantom\prime}_t J^\prime_t} \,
\delta_{J^{\vphantom\prime}_{tz} J^\prime_{tz}}
\delta_{R^{\vphantom\prime}_t R^\prime_t} \,
\delta_{I^{\vphantom\prime}_t I^\prime_t} \,
\delta_{I^{\vphantom\prime}_{tz} I^\prime_{tz}}
\delta_{Y^{\vphantom\prime}_t Y^\prime_t}
\delta_{I_t J_t} \delta_{Y_t, 0} }
\nonumber \\
& \times & (-1)^{S_{\phi^\prime} - S_{\phi^{\vphantom\prime}} +
J_\phi - J_t + (I_{B^\prime} - S_{B^\prime})
-\frac 1 2 (Y_B - \frac{N_c}{3})}
([S_B][I_{B^\prime}]
[J_{\phi^{\vphantom\prime}}] [J_{\phi^\prime}]/[I_t][J_t])^{1/2}
\nonumber \\ & \times &
\sum_{\stackrel{\scriptstyle I \in R_{\phi^{\vphantom\prime}}, \,
I^\prime \in R_{\phi^\prime} \! ,}{Y \in R_{\phi^{\vphantom\prime}}
\cap \, R_{\phi^\prime}}}
\left( \begin{array}{cc||c} R_{\phi^{\vphantom\prime}} &
R^*_{\phi^\prime} & R_t \, \gamma_t \\ I Y & I^\prime \! , \!
- \! Y & J_t \, 0 \end{array} \right)
\left( \begin{array}{cc||c} R_{\phi^{\vphantom\prime}} &
R^*_{\phi^\prime} & R_t \, \gamma_t \\ I_\phi Y_\phi
& I_{\phi^\prime}, \! - \! Y_{\phi^\prime} & I_t \, 0
\end{array} \right) \nonumber \\ & \times &
\left( \begin{array}{cc||c} R_B & R_t & R_{B^\prime} \, \tilde \gamma
\\ S_B, + \frac{N_c}{3} & J_t \, 0 & S_{B^\prime} \! , + \frac{N_c}{3}
\end{array} \right)
\left( \begin{array}{cc||c} R_B & R_t & R_{B^\prime} \, \tilde \gamma
\\ I_B Y_B & I_t \, 0 & I_{B^\prime} Y_B
\end{array} \right)
\nonumber \\ & \times & \sum_{K \tilde{K} \tilde{K}^\prime}
(-1)^{K - \frac{Y}{2}} [K] ( [\tilde{K}] [\tilde{K}^\prime] )^{1/2}
\left\{ \begin{array}{ccc}
J_\phi   & I              & K \\
I^\prime & J_{\phi^\prime} & J_t \end{array} \right\} \!
\left\{ \begin{array}{ccc}
J_\phi    & I      & K \\
\tilde{K} & S_\phi & L \end{array} \right\} \!
\left\{ \begin{array}{ccc}
J_{\phi^\prime}  & I^\prime        & K \\
\tilde{K}^\prime & S_{\phi^\prime} & L^\prime \end{array} \right\} \!
\nonumber \\ & \times &
\tau^{\left\{ I I^\prime Y \right\}}_{K \tilde{K} \tilde{K}^\prime L
L^\prime} \ , \label{tchannel}
\end{eqnarray}
\end{widetext}
where we have imposed $Y_t \! = \! 0$ but left coefficients
originating as $I_t$ or $J_t$ distinct in this expression.

[The notation used here and in Eq.~(\ref{Mmaster}) is fully explained,
among other places, in Sec.~II of Ref.~\cite{ItJt}.  For our purposes,
the most important details are: Unprimed (primed) variables represent
initial (final) quantities, $\phi$ ($B$) denote mesons (baryons),
square brackets denote representation multiplicities [$R$ for SU(3)],
and the factors with double vertical bars are SU(3) isoscalar
factors.]

$ii)$ There are also manifestly subleading terms, whose independent
reduced amplitudes multiply structures with $Y_t \! \neq \! 0$ and/or
$I_t \! \neq \! J_t$; as shown in Ref.~\cite{ItJt}, the contributions
from such amplitudes scale as $N_c^{-|Y_t|/2}$ and $N_c^{-|I_t -
J_t|}$, respectively.  One may easily transcribe an expression such as
Eq.~(\ref{tchannel}) to allow for terms with $Y_t \! \neq \!  0$ or
$I_t \! \neq \! J_t$ by inserting the appropriate $1/N_c$ power in
front of the new reduced amplitudes; this procedure has been carried
out in the 2-flavor case~\cite{CL1N,CLphoto}.

Corrections of types $i)$ and $ii)$ are dynamical in origin.  We know
of no way to obtain them without solving QCD completely.  Even if that
were feasible, such corrections would not lead to the decoupling of
spurious states unless the subleading amplitudes turned out to provide
{\em precisely\/} the factors of $-3$ necessary to cancel the
leading-order ones.  A priori it would require a truly remarkable
coincidence for this to occur.  Moreover, one would expect the values
of the subleading coefficients (being of dynamical origin) to vary
with quark masses (for illustration purposes here, taken equal and
nonzero), making decoupling for every value of quark mass even more
remarkable.  Adding to the implausibility, the reduced amplitudes
carry information not only about spurious but ordinary states as well.
Thus, if such a remarkable cancellation did occur, one would expect it
also to decouple states that must remain in the $N_c \! = \! 3$
theory.  We therefore argue that such $1/N_c$ corrections are almost
certainly not responsible for the exact decoupling of spurious $N_c \!
> \! 3$ states.

$iii)$ Unlike the 2-flavor case, the Clebsch-Gordan coefficients (CGC)
of the 3-flavor case [or more precisely, the SU(3) isoscalar factors
employed above] are not pure $O(N_c^0)$ numbers, as is the case for
2-flavor SU(2) CGC~\cite{CLSU3}.  This difference is a consequence of
the phenomenon described above, that even 3-flavor representations
containing physical states contain in addition numerous spurious
states, the subset surviving at $N_c \! = \! 3$ occupying only small
corners of the weight diagrams of these multiplets.  For transitions
between spurious and non-spurious states, the SU(3) CGC turn out to be
inhomogeneous functions of $N_c$, and couplings that are forbidden at
$N_c \! = \!  3$ are {\it all} explicitly proportional to factors of
$(1\! - \!  3/N_c)$.  In a strict large $N_c$ counting, factors such
as these are set to unity, but in considering whether states decouple
completely at $N_c \! = \! 3$ in the $1/N_c$ expansion, it is
appropriate to set such factors to zero~\cite{BL3flav}.  Consider for
example $\overline{K} \Sigma$ scattering.  The product ``{\bf
8}''$\otimes${\bf 8} contains ``{\bf 1}'' $\equiv \! [0, \frac 1 2
(N_c \! - \!  3)]$~\cite{CLSU3,CL70}.  However, for $N_c \! > \! 3$
the ``{\bf 1}'' contains not just the singlet $\Lambda$, but states
with $\Xi$ quantum numbers as well.  The SU(3) CGC for such a process,
which is finite as $N_c \! \to \! \infty$ but vanishes for $N_c \! =
\! 3$, is
\begin{equation}
\left( \begin{array}{cc||c} ``{\bf 8}\mbox{''} & {\bf 8} &
``{\bf 1}\mbox{''} \\
1, \, \frac{N_c}{3} \! - \! 1 & \frac 1 2 , \, -1 & \frac 1 2 , \,
\frac{N_c}{3} \! - \! 2 \end{array} \right)
= +\frac 1 2 \sqrt{ \frac{3 (N_c \! - \! 3)}{N_c \! + \! 5}} \ .
\end{equation}

This is the primary method of decoupling spurious states in
meson-baryon scattering.  Note the smoothness of decoupling as $N_c \!
\rightarrow \! 3$; the nature of SU(3) CGC provides a continuous path
to decoupling.  In the notation of Ref.~\cite{CLSU3penta}, the
surviving intermediate states are ${\cal N}^*$ (nonexotic) and ${\cal
E}^*_0$ [an exotic state, such as a pentaquark, not allowed for $qqq$
but still in an allowed $N_c \! = \!  3$ SU(3) representation], but
not ${\cal E}^*_1$ (spurious: only allowed for $N_c \!  > \! 3$).

Additional examples serve to illustrate this effect, and we refer to
Ref.~\cite{CLSU3} for the necessary background.  In the product ``{\bf
8}''$\otimes \,${\bf 8} for $N_c \! > \! 3$ one finds the SU(3)
representation ``{\bf S}''$\, \equiv \! [2, \frac 1 2 (N_c \! - \!
5)]$ (so named because its row of highest hypercharge carries $\Sigma$
quantum numbers), and in ``{\bf 10}''$\, \otimes \,${\bf 8} (``{\bf
10}''$\, \equiv \!  [3, \frac 1 2 (N_c \! - \! 3)]$) for $N_c \! > \!
3$ one finds an additional ``{\bf 10}'' representation, denoted ``{\bf
10}$_1$'' (the ``{\bf 10}$_2$'' defined as the one that survives at
$N_c \!  = \! 3$).  The tables of Ref.~\cite{CLSU3} provide the means
to obtain a number of their CGC via unitarity when only one product
representation has been omitted for given values of $Y$ and $I$; the
normalization sign is not determined using this method, but it can be
fixed using the methods outlined in Ref.~\cite{CLSU3} and is in any
case irrelevant for our current purposes.  In the case of
$\overline{K}N \! \to \! \Sigma$, a relevant CGC that survives the
$N_c \! \to \! \infty$ limit but vanishes for $N_c \! = \! 3$ is
\begin{equation}
\left( \begin{array}{cc||c} ``{\bf 8}\mbox{''} & {\bf 8} &
``{\bf S}\mbox{''} \\
\frac 1 2, \, \frac{N_c}{3} & \frac 1 2 , \, -1 & 1 , \,
\frac{N_c}{3} \! - \! 1 \end{array} \right)
= \pm \sqrt{ \frac{(N_c \! + \! 3)(N_c \! - \! 3)}
{(N_c \! + \! 5)(N_c \! + \! 1)}} \ ,
\end{equation}
while for the $\pi \Delta \! \to \! \Delta$ channel one encounters the
CGC
\begin{equation}
\left( \begin{array}{cc||c} ``{\bf 10}\mbox{''} & {\bf 8} &
``{\bf 10}_1\mbox{''} \\
\frac 3 2, \, \frac{N_c}{3} & 1 , \, 0 & \frac 3 2 , \,
\frac{N_c}{3} \end{array} \right)
= \pm \sqrt{ \frac{5(N_c \! + \! 5)(N_c \! - \! 3)}
{2(3N_c^2 \! + \! 14N_c \! - \! 9)}} \ .
\end{equation}

Note that this behavior is generic.  Had we lived in an $N_c \! = \!
5$ world, some of the states that are spurious at $N_c \! = \! 3$
would be physical.  However, spurious states would still occur.  The
coupling between spurious states and physical states for $N_c \! = \!
5$ all have CGC proportional to $(1 \! - \!  5/N_c)$.  Analogous
results hold for any $N_c$.

There is one other set of constraints, which originates from
recognizing that ordinary baryon resonances are obtained by scattering
mesons from baryons stable against strong decays.  In the $N_c \! = \!
3$ world, these are only the $J^P \! = \! {\frac 1 2}^+$ SU(3) {\bf 8}
baryons (less the $\Sigma^0$) and the $\Omega^-$ in the $J^P \! = \!
{\frac 3 2}^+$ {\bf 10}.  The other {\bf 10} resonances are unstable
because they decay primarily via a single $\pi$ to a state in the {\bf
8}.  In a true large $N_c$ world, all these states [those described in
Eq.~(\ref{mj})] belong to the generalization of the spin-flavor {\bf
56}, and we have noted that the ones with $J \! = \!  O(N_c^0)$ are
stable against strong decay.  The full ``{\bf 56}'' for $N_c$ large
consists of SU(3) multiplets with $J^P \!  = \! {\frac 1 2}^+ \! , \,
{\frac 3 2}^+ \! , \, \ldots , {\frac{N_c}{2}}^+$.  Therefore, all
resonances only reachable via scattering with the putatively stable
but exotic ${\frac 5 2}^+ \!  , \, {\frac 7 2}^+ \!  , \, \ldots$
baryons are spurious.  In the notation of Ref.~\cite{CLSU3penta}, the
initial states are labeled ${\cal E}$ (as opposed to nonexotic ${\cal
N}$) .  Thus, only ${\cal N} {\cal N}^*$ and ${\cal N} {\cal E}^*_0$
amplitudes survive our selection rules.

\section{Completely Broken SU(3) Flavor}

While we have made use of the full 3-flavor expression
Eq.~(\ref{tchannel}), one should not construe that using SU(3) CGC to
identify spurious states depends on SU(3) symmetry being exact.
Indeed, as we show in this section, it is meaningful even in the case
that flavor SU(3) is completely broken to SU(2)$\times$U(1).  One
begins by noting that the 3-flavor $s$-channel
expression~\cite{CLSU3penta} is found to hold separately for each
SU(3) representation $R_s$:\begin{widetext}
\begin{eqnarray}
\lefteqn{S_{L L^\prime S S^\prime J_s R_s \gamma_s
\gamma^\prime_s I_s
Y_s}} \nonumber \\
%& = & (-1)^{(S_B + S + L) - (S_{B^\prime} + S^\prime + L^\prime)}
& = & (-1)^{S_B - S_{B^\prime}}
([R_B][R_B^\prime][S][S^\prime])^{1/2} / [R_s]
%\nonumber \\ & & \times
\sum_{\stackrel{\scriptstyle I \in R_\phi, \; I^\prime \in
R_{\phi^\prime},}{I^{\prime\prime} \in R_s, \; Y \in R_\phi \cap
R_{\phi^\prime}}} (-1)^{I + I^\prime + Y} [I^{\prime\prime}]
\nonumber \\ & & \times \left( \begin{array}{cc||c} R_B & R_\phi
& R_s \, \gamma_s \\ S_B \frac{N_c}{3} & I Y & I^{\prime\prime}
\, Y \! \! + \! \frac{N_c}{3}
\end{array} \right)
\left( \begin{array}{cc||c} R_B & R_\phi & R_s \, \gamma_s \\ I_B
Y_B & I_\phi Y_\phi & I_s Y_s \end{array} \right) \nonumber \\
& & \times \left( \begin{array}{cc||c} R_{B^\prime} &
R_{\phi^\prime} & R_s \, \gamma^\prime_s \\ S_{B^\prime}
\frac{N_c}{3} & I^\prime Y & I^{\prime\prime} \, Y \! \! + \!
\frac{N_c}{3}
\end{array} \right)
\left( \begin{array}{cc||c} R_{B^\prime} & R_{\phi^\prime} & R_s
\, \gamma^\prime_s \\ I_{B^\prime} Y_{B^\prime} & I_{\phi^\prime}
Y_{\phi^\prime} & I_s Y_s \end{array} \right) \nonumber \\ & &
\times \sum_{K, \tilde{K} , \tilde{K}^\prime} [K]
([\tilde{K}][\tilde{K}^\prime])^{1/2}
%\nonumber \\ & & \times
\left\{ \begin{array}{ccc}
L   & I                & \tilde{K} \\
S   & S_B              & S_\phi    \\
J_s & I^{\prime\prime} & K \end{array} \right\} \! \left\{
\begin{array}{ccc}
L^\prime & I^\prime         & \tilde{K}^\prime \\
S^\prime & S_{B^\prime}     & S_{\phi^\prime} \\
J_s      & I^{\prime\prime} & K \end{array} \right\}
\tau^{\left\{ I I^\prime Y \right\}}_{K \tilde{K}
\tilde{K}^\prime \! L L^\prime} \ .
%\nonumber \\
\label{Mmaster}
\end{eqnarray}
\end{widetext}
Even if the poles in the scattering amplitudes do not fall into SU(3)
representations $R_s$ [completely broken SU(3)], as a matter of pure
mathematics it remains possible to separate $s$-channel contributions
according to the flavor symmetry properties of the amplitude.  The
full physical amplitude, still containing SU(3) CGC, assumes the form
\begin{widetext}
\begin{equation}
{\cal M}_{L L^\prime S S^\prime J_s R_s \gamma_s \gamma^\prime_s
I_s Y_s} = \sum_{R_s , \gamma_s} f(R_s ; m_s) S_{L L^\prime S
S^\prime J_s R_s \gamma_s \gamma^\prime_s I_s Y_s} \ ,
\end{equation}
\end{widetext}
where the coefficient functions $f(R_s ; m_s)$ equal unity in the
SU(3) limit $m_s \! \to \! m_{u,d}$ (and do not depend upon $\gamma_s$
because no physical measurement can distinguish values of this
degeneracy quantum number).

We note that the proof presented in Ref.~\cite{CL70}, that the
3-flavor expression reduces to the well-known 2-flavor
expression~\cite{MP}, uses an SU(3) CGC completeness relation
(Eq.~(13) in \cite{CL70}); in the current language, that proof
implicitly uses $f \! = \! 1$.  However, we have found the proof to
work equally well in the large $N_c$ limit regardless of the value of
$f$.  In fact, all the steps required are precisely those used to
prove the $I_t \! = \! J_t$ rule in Ref.~\cite{ItJt}.  Starting with
the expression relevant to nonstrange initial states (and similarly
for the final states):
\begin{widetext}
\begin{equation}
\sum_{R_s , \gamma_s , Y} f(R_s ; m_s) \left( \begin{array}{cc||c}
R_B & R_\phi & R_s \gamma_s \\ S_B, \frac{N_c}{3} & I Y &
I^{\prime\prime}, Y \! + \! \frac{N_c}{3}
\end{array} \right)
\left( \begin{array}{cc||c} R_B & R_\phi & R_s \gamma_s \\ S_B,
\frac{N_c}{3} & I_\phi , 0 & I_s , \frac{N_c}{3} \end{array}
\right) \ ,
\end{equation}
\end{widetext}
we note that the SU(3) completeness relation can reduce the CGC to
$\delta_{I I_\phi} \delta_{I^{\prime\prime} I_s}$ (which is precisely
what is needed to prove the result) if one can show that, at leading
order in $N_c$, only one $R_s$ and one $Y$ is allowed in the given
CGC, so that $f$ can effectively be pulled out of the sum, making its
precise value is irrelevant.  But this is precisely what is shown for
an equivalent case, Eq.~(3) of \cite{ItJt}.  The 2-flavor result
therefore follows from the 3-flavor expression, even for $f \! \neq \!
1$, i.e., for an arbitrary value of $m_s$.

\section{About $K$ Poles and Decoupling}

We have noted since our earliest papers on the subject~\cite{CL1st}
that the poles appearing in a given meson-baryon scattering amplitude
often exceed the number found in a naive quark potential model.  For
example, we found that three poles, corresponding to $K \! = \! 0$, 1,
and 2, explain the nonstrange members of the SU(6)$\times$O(3) (``{\bf
70}'',$1^-$).  For $N_c \! \ge \! 5$, this multiplet boasts two
$\Delta_{\frac 1 2}$ states, which is precisely the number of reduced
amplitudes found in the $I \! = \! \frac 3 2$, $J \! = \!
\frac 1 2$ channels ($K \! = \! 1$ and 2).  The matching of such
SU(6)$\times$O(3) multiplets to collections of resonances arising from
poles with a given $K$ was denoted ``compatibility''~\cite{CLcompat},
and was suggested as a possible explanation for the similarity of
physically observed to quark-model predicted baryon resonance
multiplets.

In the case of $N_c \! = \! 3$, however, only one state with
$\Delta_{\frac 1 2}$ quantum numbers survives in the (``{\bf
70}'',$1^-$).  Does this mean that, as one dials $N_c$ from large
values down to 3, either the $K \! = \!  1$ or $K \! = \! 2$ pole
decouples from the theory?  Clearly, such poles are not excluded on
the basis of group theory alone, because the presence or absence of
poles is a dynamical effect, as discussed in Sec.~\ref{idea}.  But
then the concept of compatibility appears to falter somewhat for small
finite $N_c$, since SU(6)$\otimes$O(3) multiplets no longer match
resonance multiplets labeled by $K$.

Indeed, this should not trouble us, for SU(6)$\otimes$O(3) is not a
true symmetry of baryon resonances, owing to the presence of large
[$O(N_c^0)$] configuration mixing~\cite{CLconfig}.  In the case of the
$\Delta_{\frac 1 2}$, the (``{\bf 70}'', $1^-$) state mixes with,
e.g., a (``{\bf 70}'', $3^-$) state.  In the limit of heavy quarks
(and large $N_c$) these states become narrow and split in mass, but
for physical values all possible mixed states should appear.  The
upshot is that in a given process poles labeled by $K$ might be
dynamically suppressed or shifted in mass, but in principle all of
them appear for finite $N_c$.

This result begs the question of how to use $N_c \! = \! 3$
phenomenology to identify which $K$ pole appears in a given process
when some might be dynamically suppressed.  As it turns out, the mixed
partial-wave amplitudes ($L \! \neq L^\prime$) for $\pi N \! \to \!
\pi \Delta$ prove very convenient for this purpose: As may be
determined from Eq.~(\ref{tchannel}), for a given value of parity at
most one value of $K$ appears in each such amplitude: $K \! = \! \frac
1 2 (L \! + \!  L^\prime) \! = \! J \! \pm \! \frac 1 2$, and $P \! =
\! (-1)^K$.  As an illustration, we note that the light
negative-parity $N_{\frac 1 2}$ states $N(1535)$ and $N(1650)$, which
we previously found from their $\pi$ and $\eta$ couplings to be mostly
$K \! = \! 0$ and 1, respectively~\cite{CL1st}, have very different
$\pi N \! \to \! \pi \Delta$ branching fractions through the $SD_{11}$
mixed partial wave ($K \! = \! 1$)~\cite{PDG}.  For $N(1535)$ it is $<
1\%$, and for $N(1650)$ it is 1--7\%, confirming the previous
assignment.

\section{Conclusions} \label{concl}

We have shown that the properties of the SU(3) Clebsch-Gordan
coefficients ensure that spurious baryon states arising in large $N_c$
QCD decouple from the physical ones surviving at $N_c \! = \! 3$.
Moreover, this decoupling is smooth; all couplings between physical
and spurious states have factors proportional to $1 \! - \! 3/N_c$.
This critical result allows the coupling to spurious states to be
treated in a self-consistent way with other $1/N_c$ corrections.  We
have furthermore seen that employing the SU(3) factors does not
require full SU(3) symmetry, and that resonant poles representing
states lying outside the naive quark model can easily occur in our
$N_c \! = \! 3$ world.

{\it Acknowledgments.}  T.D.C.\ was supported by the D.O.E.\ through
grant DE-FGO2-93ER-40762; R.F.L.\ was supported by the N.S.F.\ through
grant PHY-0456520.

\end{document}